\documentclass[12pt,a4paper]{article}
\usepackage[T1]{fontenc}
\usepackage[dvipdfmx]{graphicx}
\usepackage{bm,caption,latexsym,amssymb,amsfonts,mathtools,here}
\usepackage{bm,latexsym,amsfonts,mathtools}
\usepackage{amssymb, amsmath, comment}
\usepackage{color}
\usepackage{hyperref}
\usepackage{booktabs}
\usepackage{mathrsfs}
\usepackage{wrapfig}
\usepackage{setspace}
\numberwithin{equation}{section}
\renewcommand{\baselinestretch}{1.5}

\usepackage[sorting=none,doi=false]{biblatex}

\bibliography{reference}

\begin{document}
\begin{titlepage}
\unitlength = 1mm
\begin{flushright}
KOBE-COSMO-22-17
\end{flushright}

\vskip 1cm
\begin{center}

{ \large \textbf{  Impact of quantum entanglement induced by magnetic fields\\  on  primordial gravitational waves 
}}
\vspace{1.8cm}
\\
Sugumi Kanno$^*$, Ann Mukuno$^{\flat}$, Jiro Soda$^{\flat}$, and Kazushige Ueda$^*$
\vspace{1cm}

\shortstack[l]
{\it $^*$ Department of Physics, Kyushu University, 744 Motooka, Nishi-ku, Fukuoka 819-0395, Japan \\ 
\it $^\flat$ Department of Physics, Kobe University, Kobe 657-8501, Japan
}

\vskip 4.0cm

{\large Abstract}\\
\end{center}

There exist observational evidence to believe the existence of primordial magnetic fields generated during inflation.
We study primordial gravitational waves (PGWs) during inflation in the presence of magnetic fields sustained by a gauge kinetic coupling. In the model, not only gravitons as excitations of PGWs, but also photons as excitations of electromagnetic fields are highly squeezed.
They become entangled with each other through graviton to photon conversion and vice versa. 
We derive the reduced density matrix for the gravitons and calculate their entanglement entropy. It turns out that the state of the gravitons is not a squeezed state but a mixed state. 
Although witnessing such an entanglement is not feasible at present, it would be an important experimental challenge.

\vspace{1.0cm}
\end{titlepage}

\hrule height 0.075mm depth 0.075mm width 165mm
\tableofcontents
\vspace{1.0cm}
\hrule height 0.075mm depth 0.075mm width 165mm
\section{Introduction}

The inflationary cosmology predicts the existence of primordial gravitational waves (PGWs) that stems from quantum fluctuations. Hence, the detection of PGWs gives strong evidence of inflationary cosmology. In particular, if we could observe the quantum nature of PGWs, it would imply a discovery of gravitons. Therefore, there are several experimental projects for detecting PGWs~\cite{Hazumi:2019lys,LiteBIRD:2022cnt,Kawamura:2011zz, Amaro-Seoane:2012aqc}.
Remarkably, the quantum state of gravitons gets squeezed during inflation~\cite{Grishchuk:1989ss,Grishchuk:1990bj,Albrecht:1992kf,Polarski:1995jg,Kanno:2021vwu}.
Hence, one way to prove the quantum nature of PGWs would be to find evidence of the squeezed state of gravitons. However, it has still been a challenge to detect PGWs by laser interferometers through the statistical property of the squeezed state~\cite{Allen:1997ad,Allen:1999xw}. Some other ideas for detecting non-classical PGWs using their squeezed state are proposed.
One is to use the Hanbury Brown-Twiss interferometry, which can distinguish non-classical particles from classical ones by measuring intensity-intensity correlations~\cite{Giovannini:2010xg,Kanno:2018cuk}.
Another idea is to detect primordial gravitons indirectly by measuring their noise in the interferometers~\cite{Parikh:2020nrd,Kanno:2020usf,Parikh:2020kfh,Parikh:2020fhy} or by measuring the decoherence time of a quantum object caused by the surrounding primordial gravitons~\cite{Kanno:2021gpt}. 

We expect that the gravitons that went through inflation keep their squeezed states until today unless the environmental effects on them are considered. However, if the gravitons were surrounded by matter fields during inflation, they may not be able to keep their squeezed states anymore. We can think of a scalar field (inflaton) and a vector field as the matter field during inflation. Since the inflaton field couples with PGWs in the form of the gradient, the coupling with the vector field would be more effective. From the point of view of observations, primordial magnetic fields may have existed during inflation. In fact, there are observations that cannot be explained without the presence of primordial magnetic fields~\cite{Grasso:2000wj,Ando:2010rb,Neronov:2010gir,Essey:2010nd,Tavecchio:2010ja,Finke:2015ona}. Furthermore, if a coherence length of magnetic fields in mega-parsec scales were found, we need to consider magnetic fields generated during inflation~\cite{Turner:1987bw,Ratra:1991bn}. Hence, in this paper, we investigate whether  gravitons surrounded by  primordial magnetic fields can keep their squeezed states.

The presence of background magnetic fields causes the conversion of gravitons into photons and vice versa~\cite{Gertsenshtein:1962,Raffelt:1987im}.
These photons could be the dark photon~\cite{Masaki:2018eut}.
Therefore, we need to investigate the effect of the conversion process
on the squeezed state of gravitons. 
In our previous paper, we assumed the presence of primordial magnetic fields at the beginning of inflation and examined the evolution of the squeezing parameters of gravitons and photons in the process of graviton to photon conversion mediated by the background magnetic field~\cite{Kanno:2022ykw}.
There, it turned out that the squeezing of gravitons was robust against the conversion process. This was because the background magnetic field rapidly decays due to inflation. Then we concluded that gravitons keep their squeezed states even in the presence of the background magnetic fields.
However, if  magnetic fields decay slowly during inflation, gravitons may lose their squeezed states. 
In this case, graviton to photon conversion never ends as long as  magnetic fields survive during inflation. Hence, in this paper, we study the conversion process of gravitons in the presence of magnetic fields that decays slowly during inflation and see if the gravitons can keep their squeezed states until today.
Remarkably, we find that the magnetic fields generate the maximal entanglement between gravitons and photons. 
As a consequence, the quantum state of gravitons becomes a mixed state instead of the squeezed (pure) state. 
We note that it is a challenge to detect such entanglement directly. However, it would 
be worth pursuing the endeavor to find evidence of the quantum nature of primordial GWs.

%If the authors cannot justify this completely, then they should mention it clearly
%that we cannot witness it, and this is more of an academic value. If this is the
%case, the authors should give an honest assessment and say that witnessing such an
%entanglement is not feasible in the abstract, introduction, and in the conclusion.

The organization of the paper is as follows. In section 2, we introduce a
model describing the situation where magnetic fields are persistently generated. Then, we review the graviton-photon conversion during inflation. 
In section 3, we solve the dynamics and calculate Bogoliubov coefficients describing the time evolution of the quantum state.
We obtain four-mode squeezed state as a consequence of graviton to photon conversion.
In section 4, we calculate entanglement entropy between gravitons and photons. We discuss the quantum state at present in the presence of the entanglement.

\section{Graviton-photon conversion }

We begin with the Einstein-Hilbert action and the action for a $U(1)$ gauge field coupled to a scalar field:
\begin{eqnarray}
S=S_g+S_\phi+S_A
=\int d^4x \sqrt{-g}\,
\left[
\frac{M_{\rm pl}^2}{2}
\,R
-\frac{1}{2}(\partial_\mu \phi)(\partial^\mu \phi)-V(\phi)
-\frac{1}{4}
f^2(\phi)
F^{\mu\nu}
F_{\mu\nu}
\right]
\label{original action}\,,
\end{eqnarray}
where $M_{\rm pl}=1/\sqrt{8\pi G}$ is the Planck mass. 
The gauge field $A_\mu$ represents photons and the field strength is defined by $F_{\mu\nu}=\partial_\mu A_{\nu}-\partial_\nu A_{\mu}$.

The background inflationary dynamics is determined
by the metric 
\begin{eqnarray}
ds^2=a^2(\eta)\left[-d\eta^2+\delta_{ij} dx^idx^j\right]\,,
\end{eqnarray}
 and the inflaton $\phi (\eta)$. Once the background is given,
 the coupling function can be regarded as a function of the conformal time $\eta$;
$
f =  f(\eta) \ .
$
We also assume the presence of constant magnetic fields
$
 B_i =  {\rm constant} \,.
$
It should be emphasized that the physical magnetic fields
are not $B_i$ but $f B_i$. 
In the next section, we consider the quantum evolution of 
gravitons and photons in the above background.

\subsection{Primordial gravitational waves}
We consider gravitons in a spatially flat expanding background represented by tensor mode perturbations in the three-dimensional metric $h_{ij}$, 
\begin{eqnarray}
ds^2=a^2(\eta)\left[-d\eta^2+\left(\delta_{ij}+h_{ij}\right)dx^idx^j\right]\,,
\end{eqnarray}
where $h_{ij}$ satisfies the transverse traceless conditions $h_{ij}{}^{,j}=h^i{}_i=0$. The spatial indices $i,j,k,\cdots$ are raised and lowered by $\delta^{ij}$ and $\delta_{k\ell}$. In the case of de Sitter space, the scale factor is given by $a(\eta)=-1/(H\eta)$ where $-\infty<\eta<0$.

Expanding the Einstein-Hilbert action up to the second order in perturbations $h_{ij}$, we have
\begin{eqnarray}
\delta S_g=\frac{M_{\rm pl}^2}{8}\int d^4x\,a^2\left[
h^{ij\prime}\,h_{ij}^\prime-h^{ij,k}h_{ij,k}
\right]\,,
\label{action:g}
\end{eqnarray}
where a prime denotes the derivative with respect to the conformal time.
At this quadratic order of the action, it is convenient to expand $h_{ij}(\eta,x^i)$  in Fourier modes,
\begin{eqnarray}
h_{ij}(\eta,x^i)=\frac{2}{M_{\rm pl}} \sum_{P}\frac{1}{(2\pi)^{3/2}} \int d^3 k\,h^{P}_{\bm k}(\eta)\, e_{ij}^{P}(\bm{k})\,e^{i\bm{k}\cdot\bm{x}} 
\ ,
\label{fourier_h}
\end{eqnarray}
where three-vectors are denoted by bold math type and  $e_{ij}^{P}(\bm{k})$  are the polarization tensors  for the ${\bm k}$ mode  normalized as $e^{ijP}(\bm{k})e_{ij}^{Q}(\bm{k})=\delta^{PQ}$ with $P,Q=+,\times$. Then the action (\ref{action:g}) in the Fourier modes becomes
\begin{eqnarray}
\delta S_g=\frac{1}{2}\sum_{P}\int d^3k\,d\eta\,a^2\left[\,
|h_{\bm k}^{P\prime}|^2-k^2|h_{\bm k}^P|^2
\,\right]\,.
\label{action_fourier_h}
\end{eqnarray}

\subsection{Primordial magnetic fields}

Next, we consider the action for the photon up to the second order in perturbations $A_i$, which is given by
\begin{eqnarray}
\delta S_A=\frac{1}{2}\int d^4x\ f^2 \left[A_i^{\prime\, 2}-A_{k,i}^2\right]\,,
\label{action:A}
\end{eqnarray}
where the photon field satisfies the Coulomb gauge $A_0=0$ and $A^i{}_{,i}=0$.

If we expand the $A_i(\eta,x^i)$ in the Fourier modes, we find
\begin{align}
A_i(\eta,x^i)=\sum_{P} \frac{\pm i}{(2\pi)^{3/2}}
\int d^3 k\,A^{P}_{\bm k}(\eta)\,e_i^{P}(\bm{k})\, e^{i\bm{k}\cdot\bm{x}}
\label{fourier_A}
\ ,
\end{align}
where  $e_i^{P}(\bm{k})$ are the polarization  vectors for the ${\bm k}$ mode  normalized as $e^{iP}(\bm{k}) e_i^{Q}(\bm{k})=\delta^{PQ}$ with $P,Q=+,\times$. The sign of $\pm i$ corresponds to the $P,Q=+,\times$.
The action (\ref{action:A}) in the Fourier modes is
\begin{eqnarray}
\delta S_A=\frac{1}{2}\sum_{P}\int d^3k\,d\eta\,f^2\left[\,
|A_{\bm k}^{P\prime}|^2-k^2|A_{\bm k}^P|^2
\,\right]\,.
\label{action_fourier_A}
\end{eqnarray}

\subsection{Graviton-photon interaction}
The action for the interaction between the graviton and the photon  up to second order in perturbations $h_{ij}, A^i$ is found to be
\begin{eqnarray}
\delta S_{\rm I}=\int d^4x \left[
\varepsilon_{i\ell m} f^2 B_m h^{ij}\left(\partial_j A_\ell
-\partial_\ell A_j\right)
\right]\,.
\label{action:I}
\end{eqnarray}
Note that $B_m=\varepsilon_{mj\ell}\,\partial_j A_\ell$ is a constant background magnetic field that we assumed the presence at the beginning of inflation.

\begin{figure}[H]
\centering
 \includegraphics[keepaspectratio, scale=0.55]{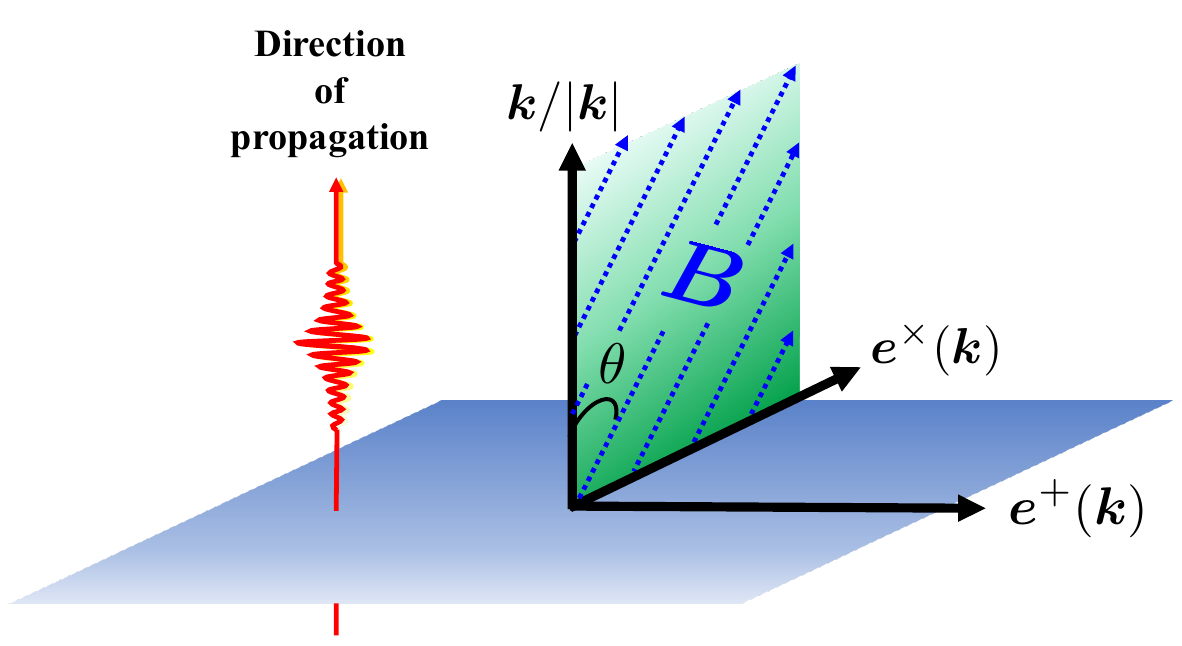} \renewcommand{\baselinestretch}{3}
 \caption{Configuration of the polarization vector ${\bm e}^P(\bm k)$, wave number ${\bm k}$, and background magnetic field ${\bm B}$.}
 \label{Configuration}
 \end{figure}

In the Fourier mode defined in Eqs.~(\ref{fourier_h}) and (\ref{fourier_A}), 
\begin{eqnarray}
\delta S_I = \frac{2}{M_{\rm pl}}\sum_{P,Q}\int d^3 k\,d\eta\,f^2\left[
\varepsilon_{i\ell m}\,B_m\,h_{\bm k}^PA_{-\bm k}^Q
\,e_{ij}^P(\bm k)\Bigl\{ik_\ell\,e_{j}^Q(-\bm k)-ik_j\,e_{\ell }^Q(-\bm k)\Bigr\}\right]
\label{action:I2}\,,
\end{eqnarray}
where $k=|\bm k|$. Polarization vectors $e^{i+}, e^{i\times}$ and a vector $k^i/k$ constitute an orthonormal basis.
Without loss of generality, we assume the constant background magnetic field is in the ($k^i, e^{i \times}$)-plane as depicted in FIG.~\ref{Configuration}.
\noindent
The polarization tensors  can be written in terms of polarization vectors  
$e^{i+}$ and $e^{i\times}$
 as
\begin{align}
    &e_{ij}^+(\bm{k})
    =\frac{1}{\sqrt{2}} \Bigl\{
    e^+_i(\bm{k}) e^+_j(\bm{k})-e^\times_i(\bm{k}) e^\times_j(\bm{k})
    \Bigr\}\,,\\
    &e_{ij}^\times(\bm{k})
    =\frac{1}{\sqrt{2}} 
    \Bigl\{
    e^+_i(\bm{k}) e^\times_j(\bm{k})+e^\times_i(\bm{k}) e^+_j(\bm{k})
    \Bigr\}\, .
\end{align}
Below, we assume
$
e_i^\times(-\bm{k})=-e_i^\times(\bm{k})
$.
The action (\ref{action:I2}) is then written as
\begin{eqnarray}
\delta S_I&=&\int d^3k\,d\eta\,f^2\,
\lambda(\bm k) \left[\,h_{\bm k}^+(\eta)\,A_{-\bm k}^+(\eta)+\,h_{\bm k}^\times(\eta)\,A_{-\bm k}^\times(\eta)\,\right]\, ,
\label{action:I3}
\end{eqnarray}
where we defined the coupling between graviton and photon as 
\begin{align}
\lambda(\bm{k})
\equiv
\frac{\sqrt{2}}{M_{\rm pl}}
\varepsilon^{i\ell m}\,e_i^+\,k_\ell\,B_m\,.
\label{coupling}
\end{align}
Here, the conditions for the graviton and photon to be real read,
$h_{-\bm k}^{+,\times}(\eta)=h_{\bm k}^{*\,+,\times}(\eta)$ and $A_{-\bm k}^{+,\times}(\eta)=-A_{\bm k}^{*\,+,\times}(\eta)$\,.
In the following, we focus on the plus polarization and omit the index $P$ unless there may be any confusion. 

\subsection{Total action in canonical variables}

If we use the canonical variable $y^P_{\bm k}(\eta)=a\,h_{\bm k}^{P}(\eta)$ and $x^P_{\bm k}(\eta)=f\,A_{\bm k}^{P}(\eta)$, the total action of Eqs.~(\ref{action_fourier_h}), (\ref{action_fourier_A}) and (\ref{action:I3}) are written as
\begin{eqnarray}
\delta S&=&\delta S_y+\delta S_x+\delta S_I
\nonumber\\
&=&\frac{1}{2}\int d^3 k\,d\eta\left[\,
|y_{\bm k}^{\prime}|^2
-\left(k^2-\left(\frac{a^\prime}{a}\right)^2\right)|y_{\bm k}|^2
-\frac{a^\prime}{a}\left(
y_{\bm k}\,y_{-\bm k}^{\prime}
+y_{-\bm k}\,y_{\bm k}^{\prime}
\right)
\right]\nonumber\\
&&+\frac{1}{2}\int d^3 k\,d\eta\left[\,
|x_{\bm k}^{\prime}|^2
-\left(k^2-\left(\frac{f^\prime}{f}\right)^2\right)|x_{\bm k}|^2
-\frac{f^\prime}{f}\left(
x_{\bm k}\,x_{-\bm k}^{\prime}
+x_{-\bm k}\,x_{\bm k}^{\prime}
\right)
\right]\,\nonumber\\
&&+\int d^3 k\,d\eta\left[\,\frac{f}{a}\,\lambda(\bm{k})\,
y_{\bm k}\,x_{-\bm k}
\,\right]\,.
\label{totalaction}  
\end{eqnarray}
The variation of the actions (\ref{totalaction}) with respect to the graviton and the photon fields gives
\begin{align}
    &y_{\bm{k}}''+\left(k^2-\frac{a^{\prime\prime}}{a}\right)y_{\bm{k}}=-\lambda f \frac{x_{\bm k}}{a} ,
    \label{eom:graviton}
    \\
    &x_{\bm{k}}''+\left(k^2-\frac{a^{\prime\prime}}{a}\right)x_{\bm{k}}=-\lambda f \frac{y_{\bm k}}{a} .
    \label{eom:photon}
\end{align}
We assume the gauge kinetic function in the form 
\begin{eqnarray}
  f(\eta)
  =a(\eta)^{-2c},
  \label{f}
\end{eqnarray}
where $c$ is a constant parameter.
We take $c=-1/2$ to make the analysis easier. For this parameter, the power spectrum of the electromagnetic fields $A_\mu$ is scale-invariant~\cite{Kanno:2009ei}. 
We have a choice of taking $c=-1$ in order to have a scale-invariant primordial magnetic field. However, we will see
$c=-1/2$ is sufficient to get
a significant modification of the quantum state of gravitons. 

\section{Time evolution of quantum state}
Using the basic equations presented in the previous section, we solve the time evolution of the quantum state in de Sitter space in this section.
We assume the effect of the coupling $\lambda$ instantaneously appears  at $\eta_i$, namely, $\lambda\,(\eta)=\lambda\,\theta(\eta-\eta_i)$.

When we define the Lagrangian in the actions (\ref{totalaction}) by $\delta S=\int d\eta\,L$, the conjugate momenta of graviton $p_{\bm k}$ and photon $\pi_{\bm k}$ are respectively given by
\begin{align}
    &p_{\bm{k}}(\eta)=\frac{\partial L}{\partial y^\prime_{-\bm k}}=y_{\bm{k}}'(\eta)+\frac{1}{\eta}y_{\bm{k}}(\eta) \ , 
    \label{p}\\
    &\pi_{\bm{k}}(\eta)=\frac{\partial L}{\partial x^\prime_{-\bm k}}=x_{\bm{k}}'(\eta)+\frac{1}{\eta}x_{\bm{k}}(\eta) \ .
    \label{pi}
\end{align}
Now we promote variables $y_{\bm k}(\eta)\,, x_{\bm k}(\eta)$ and their momenta $p_{\bm k}(\eta), \pi_{\bm k}(\eta)$ into operators. 
Annihilation operators for the graviton and photon are respectively expressed by canonical variables as
\begin{eqnarray}
&&\hat{a}_y(\eta,{\bm k})=\sqrt{\frac{k}{2}}\hat{y}_{\bm k}(\eta)+\frac{i}{\sqrt{2k}}\hat{p}_{\bm k}(\eta)\,,
\label{y:annihi}
\\
&&\hat{a}_x(\eta,{\bm k})=\sqrt{\frac{k}{2}}\hat{x}_{\bm k}(\eta)+\frac{i}{\sqrt{2k}}\hat{\pi}_{\bm k}(\eta)\,.
\label{A:annihi}
\end{eqnarray}
The commutation relations $[\hat{a}_y(\eta,{\bm k}),\hat{a}^\dag_y(\eta,-{\bm k}^\prime)]=\delta({\bm k}+{\bm k}^\prime)$ and $[\hat{a}_x(\eta,{\bm k}),\hat{a}^\dag_x(\eta,-{\bm k}^\prime)]=\delta({\bm k}+{\bm k}^\prime)$ guarantee the canonical commutation relations $[\hat y_{\bm k}(\eta), \hat p_{{\bm k}^\prime}(\eta)]=i\delta({\bm k}-{\bm k}^\prime)$ and $[\hat x_{\bm k}(\eta),\hat \pi_{{\bm k}^\prime}(\eta)]=i\delta({\bm k}-{\bm k}^\prime)$.
Note that the annihilation operator becomes time-dependent through the time dependence of canonical variables. Thus, the vacuum defined by $\hat{a}(\eta,{\bm k})|0\rangle =0$ is time dependent as well, and the vacuum in this formalism turns out to be defined at every moment.
Our aim is to find the formula for Bogoliubov coefficients
relating $\hat{a}_y(\eta,{\bm k}), \hat{a}_x(\eta,{\bm k}) $
and $\hat{a}_y(\eta_i,{\bm k}), \hat{a}_x(\eta_i,{\bm k}) $.

\subsection{Boundary conditions}
In this subsection, we specify boundary conditions of solutions of Eqs. (\ref{eom:graviton}) and (\ref{eom:photon}).
Notice that $\lambda =0$ before the initial time $\eta_i$. 

Let us first consider Eqs.~(\ref{eom:graviton}) and (\ref{eom:photon}) of the form
\begin{align}
    &\hat{y}_{\bm{k}}^{(0)\prime\prime}+\left(k^2-\frac{2}{\eta^2}\right)\hat{y}_{\bm{k}}^{(0)}=0\,,
    \label{GWeqs}
    \\
    &\hat{x}_{\bm{k}}^{(0)\prime\prime}+\left(k^2-\frac{2}{\eta^2}\right)
    \hat{x}_{\bm{k}}^{(0)}=0 \, ,
    \label{EMeqs}
\end{align}
where the superscript $(0)$ denotes $\lambda=0$. Since Eqs.~(\ref{GWeqs}) and (\ref{EMeqs}) are the same form, the mode function for the graviton and the photon at the zeroth order becomes identical. Then the solutions of the above equations can be written as
\begin{align}
    &\hat{y}^{(0)}_{\bm k}(\eta)=u_{\bm k}(\eta) ~\hat{c}
    +u_{\bm k}^*(\eta)~\hat{c}^\dagger\,, 
    \label{0th:graviton}\\
    &\hat{x}^{(0)}_{\bm k}(\eta)=u_{\bm k}(\eta) ~\hat{d}
    +u^*_{\bm k}(\eta)~\hat{d}^\dagger,
    \label{0th:photon}
\end{align}
where $\hat{c}$\,($\hat{d}$) and its conjugate 
$\hat{c}^\dag$($\hat{d}^\dag$) are constant operators of integration. We choose the properly normalized positive frequency mode in the remote past as a basis, which is expressed as
\begin{align}
    u_{\bm k}(\eta)=\frac{1}{\sqrt{2k}} \biggl(1-\frac{i}{k\eta}\biggr) e^{-ik\eta}.
    \label{u}
\end{align}
Thus, annihilation operators at the initial time are expressed by the zeroth order variables
\begin{align}
    \hat{a}_y(\eta_i,\bm{k})
    &=\left(1-\frac{i}{2k\eta_i}\right)e^{-ik\eta_i}\,\hat{c}
    +\frac{i}{2k\eta_i} e^{ik\eta_i}\,\hat{c}^\dagger,
    \label{ycRel}
    \\
    \hat{a}_x(\eta_i,\bm{k})
    &=\left(1-\frac{i}{2k\eta_i}\right)e^{-ik\eta_i}\,\hat{d}
    +\frac{i}{2k\eta_i} e^{ik\eta_i}\,\hat{d}^\dagger.
\label{AdRel}
\end{align}
Combining Eqs. (\ref{ycRel}) and (\ref{AdRel}) with their complex conjugate, we can express the  $\hat{c}$ and the  $\hat{d}$  by the initial creation and annihilation 
operators as
\begin{eqnarray}
    \hat{c}  &=& \left( 1+\frac{i}{2k\eta_i}\right)e^{ik\eta_i}\,\hat{a}_y(\eta_i,\bm{k} )
    -\frac{i}{2k\eta_i} e^{ik\eta_i}\,\hat{a}_y^\dagger (\eta_i,-\bm{k}) \ , \\
  \hat{d}  &=&   \left( 1+\frac{i}{2k\eta_i}\right)e^{ik\eta_i}\,\hat{a}_x(\eta_i,\bm{k} )
    -\frac{i}{2k\eta_i} e^{ik\eta_i}\,\hat{a}_x^\dagger (\eta_i,-\bm{k}) \ .
\end{eqnarray}
Thus, we have obtained the boundary conditions. 
By solving Eqs. (\ref{eom:graviton}) and (\ref{eom:photon}) with $f=a$ analytically, we will take into account the effect of interaction between gravitons and photons in the next subsection.

\subsection{Bogoliubov coefficients}

To properly take into account the boundary conditions at $\eta_i$, it is convenient to diagonalize equations of motion (\ref{eom:graviton}), (\ref{eom:photon}) as
\begin{align}
    &Y_{\bm{k}}''+\left(k^2-\frac{2}{\eta^2}\,+\, \lambda \right)Y_{\bm{k}}=0 ,
    \label{diagonalEOM1}\\
    &X_{\bm{k}}''+\left(k^2-\frac{2}{\eta^2}\,-\, \lambda \right)X_{\bm{k}}=0 ,
    \label{diagonalEOM2}
\end{align}
where we defined $Y_{\bm k}=(x_{\bm k}+y_{\bm k})/2$ and $X_{\bm k}=(x_{\bm k}-y_{\bm k})/2$.
The solutions of Eqs.(\ref{diagonalEOM1}) and  (\ref{diagonalEOM2}) are written as 
\begin{align}
&\tilde{Y}_{\bm k}(\eta,\lambda)=c_1\,Y_{\bm k}(\eta,\lambda) +c_2\,Y_{\bm k}^*(\eta,\lambda)~ ,\\
&\tilde{X}_{\bm k}(\eta,\lambda)=c_1\,X_{\bm k}(\eta,\lambda) +c_2\,X_{\bm k}^*(\eta,\lambda)~ ,
\end{align}
where $c_1$ and $c_2$ are constants of integration. We defined 
\begin{align}
&Y_{\bm k}(\eta,\lambda)={\frac{1}{\sqrt{2\sqrt{k^2+\lambda}}}}\left( 1-\frac{i}{\sqrt{k^2+\lambda}\,\eta} \right)e^{-i\sqrt{k^2+\lambda}\,\eta}~\,,\\
&X_{\bm k}(\eta,\lambda)=
{\frac{1}{\sqrt{2\sqrt{k^2-\lambda}}}}\left( 1-\frac{i}{\sqrt{k^2-\lambda}\,\eta} \right)e^{-i\sqrt{k^2-\lambda}\,\eta}~
\,.
\end{align}
Note that we take the same constants of integration for $\tilde{Y}_{\bm k}$ and $\tilde{X}_{\bm k}$ so that  $\tilde{Y}_{\bm k}(\eta,\lambda)$ and $\tilde{X}_{\bm k}(\eta,\lambda)$ are interchanged by switching
$\lambda$ to $-\lambda$. Then we can construct the odd and even solution with respect to $\lambda$ as 
\begin{align}
& F^{\,(\rm odd)}_{\bm k}(\eta,\lambda)=\frac{1}{2}\left( \tilde{Y}_{\bm k}(\eta,\lambda)-\tilde{X}_{\bm k}(\eta,\lambda) \right),
\label{oddsolution}
\\
&F^{\,(\rm even)}_{\bm k}(\eta,\lambda)=\frac{1}{2}\left( \tilde{Y}_{\bm k}(\eta,\lambda)+\tilde{X}_{\bm k}(\eta,\lambda) \right),
\label{evensolution}
\end{align}
respectively. Here, we call Eq.~(\ref{oddsolution}) odd solution because a minus sign comes out by switching $\lambda$ to $-\lambda$.
The coefficients $c_1$ and $c_2$ are given by the junction condition at $\eta=\eta_i$.% so that the solution is smoothly connected to the Bunch Davies vacuum at the initial time
%\begin{align}
%&F^{\,(\rm odd)}_{\bm k}(\eta_i,\lambda)=F^{\,(\rm odd)}_{\bm k}(\eta_i,0)=0~,~~F^{\,(\rm odd)\prime}_{\bm k}(\eta_i,\lambda)=F^{\,(\rm odd)\prime}_{\bm k}(\eta_i,0),\\
%&F^{\,(\rm even)}_{\bm k}(\eta_i,\lambda)=F^{\,({\rm even})}_{\bm k}(\eta_i,0)=u_{\bm k} (\eta_i)~,~~
%F^{\,({\rm even})\prime}_{\bm k}(\eta_i,\lambda)=F^{\,({\rm even})\prime}_{\bm k}(\eta_i,0) .
%\end{align}

Let us solve the dynamics with the boundary conditions
\begin{eqnarray}
\hspace{-1cm}
\hat{y}_{\bm k}(\eta,\lambda)|_{\eta<\eta_i}&=& u_{\bm k}(\eta) \ ,
\\
\label{pGWsol}
\hat{x}_{\bm k}(\eta,\lambda)|_{\eta<\eta_i} &=&u_{\bm k}(\eta)\ .
\end{eqnarray}
where we assumed that the positive frequency mode of de Sitter space is realized before the time $\eta_i$. After the time $\eta_i$, the gravitons and photons start to interact each other. 
Taking into account the relations $Y_{\bm k}=(x_{\bm k}+y_{\bm k})/2$ and $X_{\bm k}=(x_{\bm k}-y_{\bm k})/2$,
we can deduce the graviton field and its conjugate momentum as
\begin{eqnarray}
\hspace{-1cm}
\hat{y}_{\bm k}(\eta,\lambda)&=& F^{\,(\rm even)}_{\bm k}(\eta,\lambda) \,\hat{c}
+F^{\,(\rm odd)}_{\bm k}(\eta,\lambda)\,\hat{d}
+{\rm H.c.} \,,
\\
\label{pGWsol}
\hat{p}_{\bm k}(\eta,\lambda)&=&
F^{\,({\rm even})\prime}_{\bm k}
(\eta,\lambda) 
\,\hat{c}
+F^{\,({\rm odd})\prime}_{\bm k}(\eta,\lambda) 
\,\hat{d} \nonumber\\
&&~~~~~~~~~~~
+\frac{1}{\eta}\Bigl(
F^{\,({\rm even})}_{\bm k}(\eta,\lambda)
\,\hat{c}
+
F^{\,({\rm odd})}_{\bm k}(\eta,\lambda)\,\hat{d}\,\Bigr)+{\rm H.c.}\,,
\end{eqnarray}
where we used Eq.~(\ref{p}) and H.c. represents Hermitian conjugate. We see that the operator of photon $\hat{d}$ comes into the graviton $\hat{y}_{\bm k}$ and $\hat{p}_{\bm k}$ together with $F_{\bm k}^{({\rm odd})}$.
For the photon,
the field and its conjugate momentum become
\begin{eqnarray}
\hat{x}_{\bm k}(\eta)&=&
F^{\,({\rm even})}_{\bm k}(\eta,\lambda)
\,\hat{d}
+
F^{\,({\rm odd})}_{\bm k}(\eta,\lambda)
\,\hat{c}
+{\rm H.c.}\,,\\
\hat{\pi}_{\bm k}(\eta)&=&
F^{\,({\rm even})\prime}_{\bm k}(\eta,\lambda)
\,\hat{d}
+
F^{\,({\rm odd})\prime}_{\bm k}(\eta,\lambda)
\,\hat{c} \nonumber\\
&&~~~~~~~~~~~
+\frac{1}{\eta}\Bigl(
F^{\,({\rm even})}_{\bm k}(\eta,\lambda)
\,\hat{d}
+
F^{\,({\rm odd})}_{\bm k}(\eta,\lambda)
\,\hat{c}
\, \Bigr)
+{\rm H.c.}\,,
\label{pEMsol}
\end{eqnarray}
where we used Eq.~(\ref{pi}). We see that the operator of graviton $\hat{c}$ comes into the photon $\hat{x}_{\bm k}$ and $\hat{\pi}_{\bm k}$ together with $F_{\bm k}^{\,({\rm odd})}$.
Then the annihilation operators for the graviton and photon are obtained by using Eqs.~(\ref{y:annihi}) and (\ref{A:annihi}) such as
\begin{align}
    \hat{a}_{y}(\eta,\bm{k})
    &= 
    \psi_{p}^{({\rm even})}
    \, \hat{c}
    +\psi_{m}^{({\rm even})*}
    \hat{c}^\dagger
    +\psi_{p}^{({\rm odd})} \hat{d}
    +\psi_{m}^{({\rm odd})*} \hat{d}^\dagger\,,
    \label{y:annihifull}
    \\
    \hat{a}_{x}(\eta,\bm{k})
    &=\psi_{p}^{({\rm even})}\,\hat{d}
    +\psi_{ m}^{({\rm even})*}\, \hat{d}^\dagger
    +\psi_p^{({\rm odd})} \hat{c}
    +\psi_m^{({\rm odd})*} \hat{c}^\dagger\,.
    \label{A:annihifull}
\end{align}
Here, we defined new variables
\begin{align}
&\psi_p^{(j)} 
    =\sqrt{\frac{k}{2}} 
    F^{\,(j)}_{\bm k}(\eta,\lambda)
    +\frac{i}{\sqrt{2k}}
    \Bigl(
    F^{\,(j)\prime}_{\bm k}(\eta,\lambda)
    +\frac{1}{\eta}
    F^{\,(j)}_{\bm k}(\eta,\lambda)
    \Bigr) ,\label{psip}\\
&\psi_m^{(j)}
     =\sqrt{\frac{k}{2}} 
    F^{\,(j)}_{\bm k}(\eta,\lambda)
    -\frac{i}{\sqrt{2k}}
    \Bigl(
    F^{\,(j)\prime}_{\bm k}(\eta,\lambda)
    +\frac{1}{\eta}
    F^{\,(j)}_{\bm k}(\eta,\lambda)
    \Bigr), \label{psim}
\end{align}
where $(j)=({\rm even}),({\rm odd})$ denotes the even mode and 
 odd mode with respect to the coupling $\lambda$, respectively.

Inserting the above back into Eqs. (\ref{y:annihifull}) and (\ref{A:annihifull}), the time evolution of the annihilation operator of the graviton is described by the Bogoliubov transformation in the form
\begin{align}
    &\hat{a}_y(\eta,\bm{k})=
    \Biggl[
    \psi_{p}^{({\rm even})}
    \Bigl( 1+\frac{i}{2k\eta_i} \Bigr)
    e^{ik\eta_i} 
    +\psi_{m}^{({\rm even})*}
     \frac{i}{2k\eta_i} e^{-ik\eta_i} 
    \Biggr] \hat{a}_y(\eta_i,\bm{k})\nonumber\\
    &\hspace{1.5cm}
    +\Biggl[
    \psi_{p}^{({\rm even})}
    \Bigl(-\frac{i}{2k\eta_i} \Bigr) e^{ik\eta_i} 
    +
    \psi_{m}^{({\rm even})*}
    \Bigl(1-\frac{i}{2k\eta_i}\Bigr)
    e^{-ik\eta_i} \Biggr]\hat{a}_y^\dagger(\eta_i,-\bm{k})
    \nonumber\\
     &\hspace{1.5cm}
    +\Biggl[
    \psi_{p}^{({\rm odd})}
    \Bigl(1+\frac{i}{2k\eta_i} \Bigr) e^{ik\eta_i} 
    +
    \psi_{m}^{({\rm odd})*}
    \frac{i}{2k\eta_i}
    e^{-ik\eta_i} \Biggr]\hat{a}_x(\eta_i,\bm{k})
    \nonumber\\
     &\hspace{1.5cm}
    +\Biggl[
    \psi_{p}^{({\rm odd})}
    \Bigl(-\frac{i}{2k\eta_i} \Bigr) e^{ik\eta_i} 
    +\psi_m^{({\rm odd})*}
    \Bigl(1-\frac{i}{2k\eta_i}\Bigr)
    e^{-ik\eta_i} \Biggr]\hat{a}_x^\dagger(\eta_i,-\bm{k}),
    \label{y:bogoliubov1}
\end{align}
and the time evolution of annihilation operator of photon is expressed by the Bogoliubov transformation such as
\begin{align}
    &\hat{a}_x(\eta,\bm{k})=
    \Biggl[
    \psi_{p}^{({\rm even})}
 \Bigl( 1+\frac{i}{2k\eta_i} \Bigr)
    e^{ik\eta_i} 
    +\psi_{m}^{({\rm even})*}
    \frac{i}{2k\eta_i} e^{-ik\eta_i} 
    \Biggr] \,    \hat{a}_x(\eta_i,\bm{k})\nonumber\\
    &\hspace{1.5cm}
    +\Biggl[
    \psi_{p}^{({\rm even})}
    \Bigl(-\frac{i}{2k\eta_i} \Bigr) e^{ik\eta_i} 
    +\psi_{m}^{({\rm even})*}
    \Bigl(1-\frac{i}{2k\eta_i}\Bigr)
    e^{-ik\eta_i} \Biggr]\hat{a}_x^\dagger(\eta_i,-\bm{k})
    \nonumber\\
     &\hspace{1.5cm}
    +\Biggl[
    \psi_{p}^{({\rm odd})}
    \Bigl(1+\frac{i}{2k\eta_i} \Bigr) e^{ik\eta_i} 
    +
    \psi_{m}^{({\rm odd})*}
    \frac{i}{2k\eta_i}
    e^{-ik\eta_i} \Biggr]\hat{a}_y(\eta_i,\bm{k})
    \nonumber\\
     &\hspace{1.5cm}
    +\Biggl[
    \psi_{p}^{({\rm odd})}
    \Bigl(-\frac{i}{2k\eta_i} \Bigr) e^{ik\eta_i} 
    +\psi_m^{({\rm odd})*}
    \Bigl(1-\frac{i}{2k\eta_i}\Bigr)
    e^{-ik\eta_i} \Biggr]\hat{a}_y^\dagger(\eta_i,-\bm{k}).
    \label{A:bogoliubov1}
\end{align}
We see that the Bogoliubov transformations for graviton and photon are symmetric each other. The Bogoliubov transformations show the particle production during inflation and the mixing between graviton and photon.

Let us introduce a matrix form of the Bogoliubov transformations for the calculations below. The Bogoliubov transformation (\ref{y:bogoliubov1}) and (\ref{A:bogoliubov1}) and their conjugate can be accommodated into the simple $4\times 4$ matrix form $M$
\begin{eqnarray}
\begin{pmatrix}
a_y(\eta)\\
a_y^{\dagger}(\eta)\\
a_x(\eta)\\
a_x^{\dagger}(\eta)\\
\end{pmatrix}
=M
\begin{pmatrix}
a_y(\eta_i)\\
a_y^{\dagger}(\eta_i)\\
a_x(\eta_i)\\
a_x^{\dagger}(\eta_i)\\
\end{pmatrix}
=
\begin{pmatrix}
A & B\\
B  & A\\
\end{pmatrix} 
\begin{pmatrix}
a_y(\eta_i)\\
a_y^{\dagger}(\eta_i)\\
a_x(\eta_i)\\
a_x^{\dagger}(\eta_i)\\
\end{pmatrix}\ .
\hspace{-6mm}
\label{bogoliubov}
\end{eqnarray}
Here, the $M$ consists of $2\times 2$ matrices $A$ and $B$. The $A$ is written by the even-order solutions, 
\begin{eqnarray}
A_{\rm even}=
\begin{pmatrix}
K_{\rm even}^* & -L_{\rm even}^* \\
-L_{\rm even} & K_{\rm even} \\
\end{pmatrix}\,,
\end{eqnarray}
where we defined 
\begin{align}
&K_{\rm even}=\psi^{({\rm even})*}_p \Bigl(
1-\frac{i}{2 k \eta_i}
\Bigr)
e^{-ik\eta_i}
-
\psi^{({\rm even})}_m 
\frac{i}{2 k \eta_i}
e^{ik\eta_i}, \\
&L_{\rm even}=-\psi^{({\rm even})*}_p 
\frac{i}{2 k \eta_i}
e^{-ik\eta_i}
-
\psi^{({\rm even})}_m 
\Bigl(
1+\frac{i}{2 k \eta_i}
\Bigr)
e^{ik\eta_i}.
\end{align}

The $B$ comes from the first order solution expressed as
\begin{eqnarray}
B =
\begin{pmatrix}
K_{\rm odd}^* & -L_{\rm odd}^*\\
-L_{\rm odd}&K_{\rm odd}\\
\end{pmatrix}\,,
\end{eqnarray}
where we defined
\begin{eqnarray}
K_{\rm odd}&=&\psi^{({\rm odd})*}_p \Bigl(
1-\frac{i}{2 k \eta_i}
\Bigr)
e^{-ik\eta_i}
-
\psi^{({\rm odd})}_m 
\frac{i}{2 k \eta_i}
e^{ik\eta_i}, \\
L_{\rm odd}&=&-\psi^{({\rm odd})*}_p 
\frac{i}{2 k \eta_i}
e^{-ik\eta_i}
-
\psi^{({\rm odd})}_m 
\Bigl(
1+\frac{i}{2 k \eta_i}
\Bigr)
e^{ik\eta_i}.
\end{eqnarray}

\subsection{Inversion of the Bogoliubov transformation}

In the previous subsection, we obtained the Bogoliubov transformation that mix the operators $\hat{a}_y(\eta,{\bm k})$, $\hat{a}_x(\eta,{\bm k})$
and their Hermitian conjugates $\hat{a}^\dagger_y(\eta,-{\bm k})$, $\hat{a}^\dagger_x(\eta,-{\bm k})$.
Then the initial state is defined by
\begin{eqnarray}
\hat{a}_y(\eta_i,\bm{k})  |\overline{{\rm BD}}\rangle= \hat{a}_x(\eta_i,\bm{k}) |\overline{{\rm BD}}\rangle =0\,.
\label{BD}
\end{eqnarray}
Here, $|\overline{{\rm BD}}\rangle$ is  a vacuum  deviated from the Bunch-Davies vacuum due to the presence of the constant background magnetic field.
In order to impose these conditions,
we need to invert the Bogoliubov transformations (\ref{y:bogoliubov1}) and (\ref{A:bogoliubov1})
into the form
\begin{eqnarray}
\hat{a}_y(\eta_i,\bm{k})
&=&\alpha\,\hat{a}_y (\eta,\bm{k})+ \beta\,\hat{a}_y^\dagger(\eta,-\bm{k})
  +\gamma\,\hat{a}_x(\eta,\bm{k})+ \delta\,\hat{a}_x^\dagger(\eta,-\bm{k})\,,
  \label{y:invert}\\
\hat{a}_x(\eta_i,\bm{k})
&=&\gamma \,\hat{a}_y(\eta,\bm{k})+ \delta \,\hat{a}_y^\dagger(\eta,-\bm{k}) 
  +\alpha \,\hat{a}_x (\eta,\bm{k})+ \beta \,\hat{a}_x^\dagger(\eta,-\bm{k})\,,
  \label{A:invert}
\end{eqnarray}
where $\alpha$, $\beta$, $\gamma$, $\delta$ are the Bogoliubov coefficients. 
In order to find these coefficients, we need the inverse of the matrix $M$, which is calculated as 
\begin{eqnarray}
M^{-1}=
\begin{pmatrix}
\left( A - B A^{-1} B\right)^{-1} & -\left( A B^{-1}A - B\right)^{-1}\\
-\left( A B^{-1}A - B\right)^{-1}  & \left( A - B A^{-1} B\right)^{-1}\\
\end{pmatrix} 
\, .\label{inverse1}
\end{eqnarray}
From Eqs.~(\ref{y:invert}) and (\ref{A:invert}), the $M^{-1}$ is also written as
\begin{eqnarray}
M^{-1}=
\begin{pmatrix}
\alpha & \beta & \gamma & \delta \\
\beta^* & \alpha^* & \delta^* & \gamma^* \\
\gamma & \delta & \alpha & \beta \\
\delta^* & \gamma^* & \beta^* & \alpha^* \\
\end{pmatrix} \, .
\label{inverse2}
\end{eqnarray}
By comparing Eq.(\ref{inverse1}) with (\ref{inverse2}), we can obtain the Bogoliubov coefficients
$\alpha$, $\beta$, $\gamma$, $\delta$ numerically.

\subsection{Squeezed state}
In the previous subsection, we obtained the Bogoliubov coefficients of Eqs.~(\ref{y:invert}) and (\ref{A:invert}). If we apply the Eqs.~(\ref{y:invert}) and (\ref{A:invert}) to the definition of the initial state~(\ref{BD}) and use the relations $[\hat{a}_y(\eta,{\bm k}),\hat{a}^\dag_y(\eta,-{\bm k}^\prime)]=\delta({\bm k}+{\bm k}^\prime)$\,, $[\hat{a}_x(\eta,{\bm k}),\hat{a}^\dag_x(\eta,-{\bm k}^\prime)]=\delta({\bm k}+{\bm k}^\prime)$ and $[\hat{a}_y(\eta,{\bm k}),\hat{a}_x(\eta,-{\bm k}^\prime)]=0$, the initial state can be written by using squeezing parameters $\Lambda$ and $\Xi$ of the form
\begin{eqnarray}
|\overline{{\rm BD}}\rangle  = \prod_{k=-\infty}^{\infty}\exp\left[\frac{\Lambda}{2}\,
\hat{a}_y^\dag (\eta,\bm{k}) \hat{a}_y^\dag (\eta,-\bm{k})+\Xi\,\hat{a}_y^\dag (\eta,\bm{k}) \hat{a}_x^\dag (\eta,-\bm{k})
+\frac{\Lambda}{2}\,\hat{a}_x^\dag (\eta,\bm{k}) \hat{a}_x^\dag (\eta,-\bm{k})\right]|0\rangle,\nonumber
\label{BD0}
\\
\end{eqnarray}
where $|0\rangle$ is the instantaneous vacuum defined by
\begin{eqnarray}
\hat{a}_y(\eta,{\bm k})  |0\rangle=\hat{a}_x(\eta,{\bm k}) |0\rangle =0 \,.
\end{eqnarray}
Note that $\Lambda$ and $\Xi$ are complex parameters. The squeezing parameter of a graviton-graviton pair and a photon-photon pair is written by the same $\Lambda$. This is because the Bogoliubov transformations (\ref{y:bogoliubov1}) and (\ref{A:bogoliubov1}) are symmetric. The squeezing of graviton-photon pair is expressed by the $\Xi$. 
This describes a four mode squeezed state of pairs of graviton $y$ and photon $x$. In a different context, a four-mode squeezed state of two free massive scalar fields is discussed in~\cite{Albrecht:2014aga,Kanno:2015ewa}.
If we expand the exponential function in Taylor series, we find
\begin{eqnarray}
|\overline{{\rm BD}}\rangle  =
\prod_{\bm k} \sum_{p\,,q\,,r=0}^{\infty}
 \frac{\Lambda^{p+r}\,\Xi^q}{2^{p+r}p!\,q!\,r!}  
 |p+q \rangle_{y,{\bm k}} \otimes |p \rangle_{y,-{\bm k}} \otimes |r \rangle_{x,{\bm k}} \otimes |q+r \rangle_{x,-{\bm k}}\,.
\end{eqnarray}
This is a four-mode squeezed state which consists of an infinite number of entangled particles in the ${\cal H}_{y,{\bm k}}\otimes{\cal H}_{y,{-\bm k}}\otimes{\cal H}_{x,{\bm k}}\otimes{\cal H}_{x,-{\bm k}}$ space. 
In particular, in the highly squeezing limit $\Lambda\,,\Xi\rightarrow 1$, the Bunch-Davies vacuum becomes the maximally entangled state from the point of view of the instantaneous vacuum. 

Now we find the squeezing parameters $\Xi$ and $\Lambda$.
The condition $\hat{a}_y(\eta_i,{\bm k})|\overline{{\rm BD}}\rangle=0$ of Eq.~(\ref{BD}) yields
\begin{eqnarray}
\alpha \Lambda +\beta +\gamma \Xi =0 \ , \qquad
\alpha \Xi +\gamma \Lambda +\delta  =0\,,
\end{eqnarray}
and another condition $\hat{a}_x(\eta_i,{\bm k}) |\overline{{\rm BD}}\rangle=0$ gives the same equations.
Then, we obtain the two squeezing parameters $\Lambda$ and $\Xi$ of the form
\begin{eqnarray}
\Lambda= \frac{\gamma\delta -\beta \alpha}{\alpha^2 -\gamma^2} \ , \qquad
\Xi= \frac{\beta\gamma - \alpha\delta}{\alpha^2 -\gamma^2} \, .
\label{squeezingparameters}
\end{eqnarray}

\begin{figure}[H]
\centering
 \includegraphics[keepaspectratio,width=0.68\linewidth]{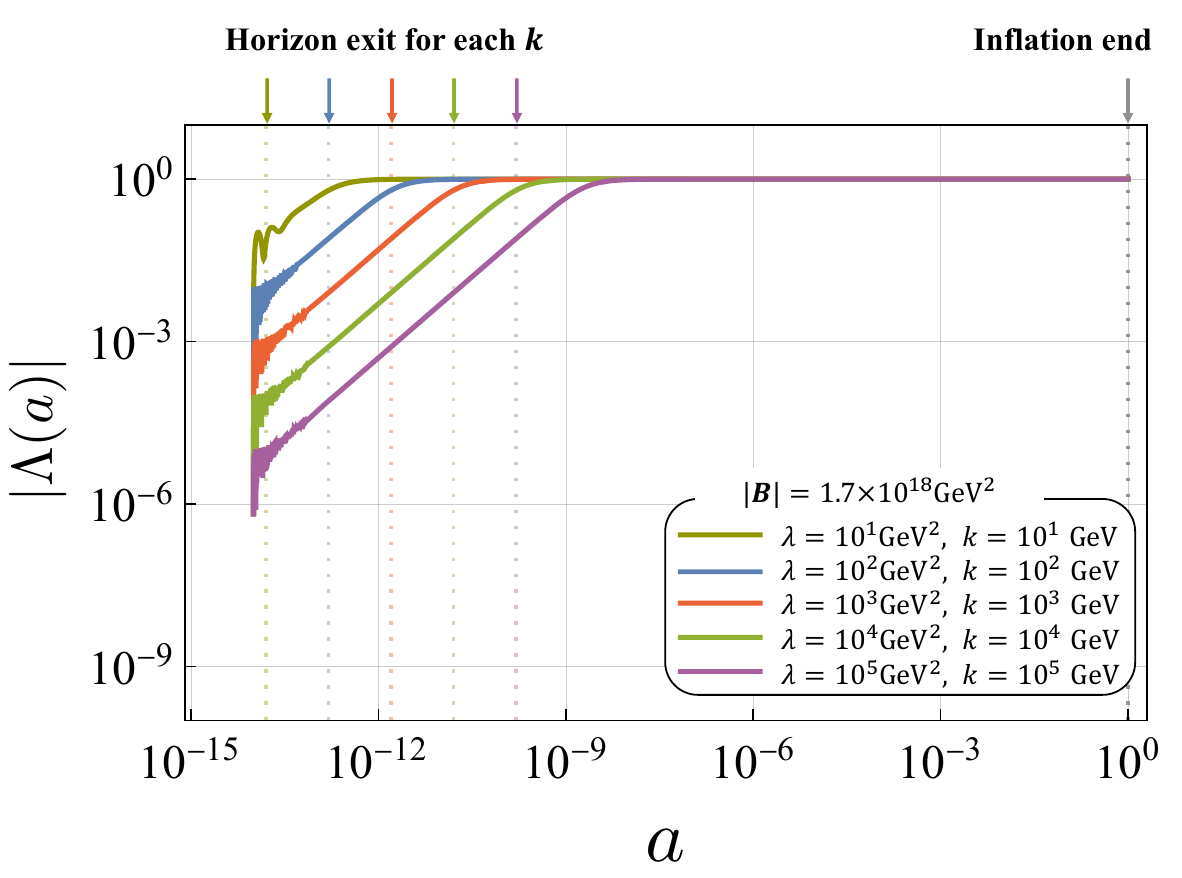}
 \renewcommand{\baselinestretch}{3}
 \caption{The squeezing parameter of gravitons or photons as a function of the scale factor of $a(\eta)$. Other parameters are set as $H=10^{14}\,{\rm GeV}, \eta_i=-1\,{\rm GeV}^{-1}, \eta_f=-10^{-14}\,{\rm GeV}^{-1}, a(\eta_i)=10^{-14},a(\eta_f)=1$.}
 \label{graph1}
 \end{figure}

\begin{figure}[H]
\centering
 \includegraphics[keepaspectratio,width=0.68\linewidth]{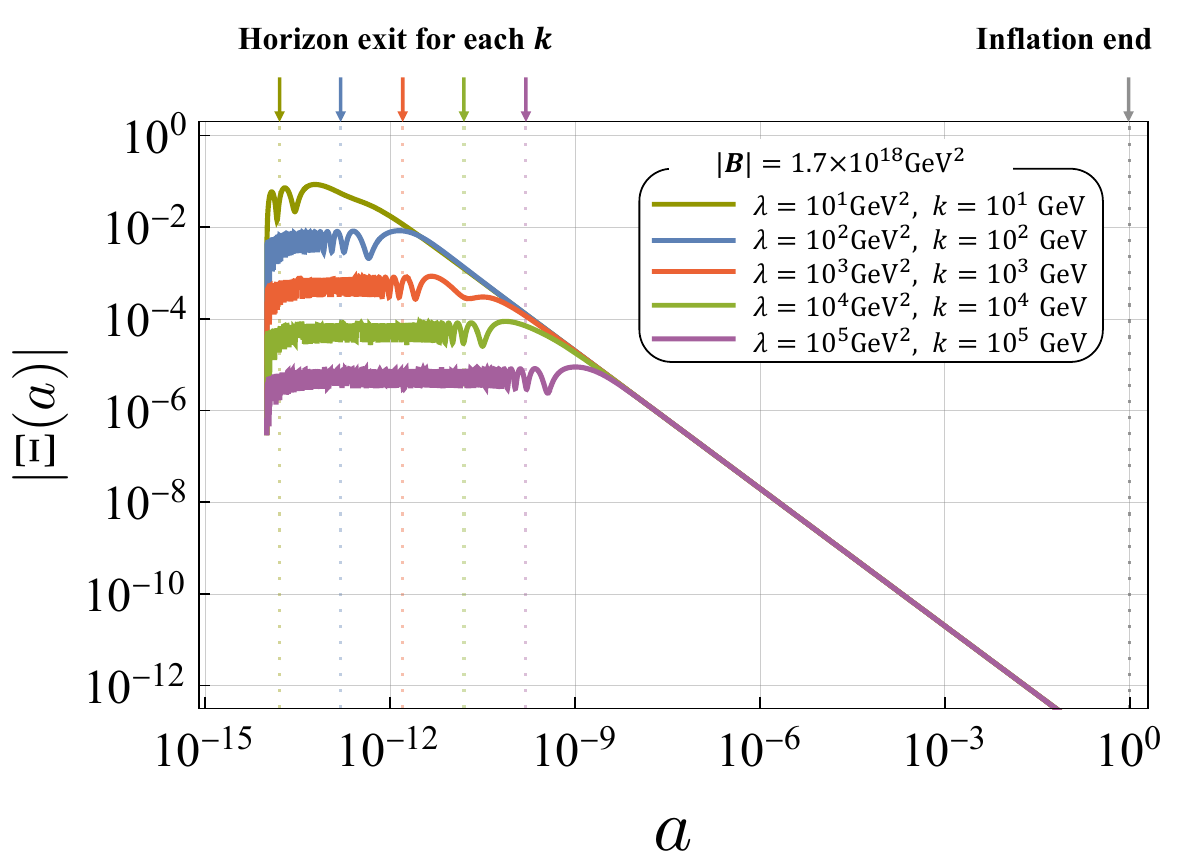}
 \renewcommand{\baselinestretch}{3}
 \caption{The squeezing parameter of graviton and photon pair as a function of $a(\eta)$. Other parameters are set as $H=10^{14}\,{\rm GeV}, \eta_i=-1\,{\rm GeV}^{-1}, \eta_f=-10^{-14}\,{\rm GeV}^{-1}, a(\eta_i)=10^{-14},a(\eta_f)=1$.}
 \label{graph2}
 \end{figure}

The results of numerical calculations for the squeezing parameters $\Lambda$ and $\Xi$ versus $a(\eta)$ with  different values of $k$ are plotted in FIGs.~\ref{graph1} and \ref{graph2}, respectively, where we normalized the scale factor at the end of the inflation as $a(\eta_f)=1$. We note that the coupling Eq.~{\ref{coupling}} is expressed as
\begin{align}
\lambda(\bm{k})
\equiv
\frac{\sqrt{2}}{M_{\rm pl}}
k\,|{\bf B}| \sin\theta \, ,
\label{coupling2}
\end{align}
where $\theta$ is the angle between the magnetic field and the wave number vectors depicted in FIG.~\ref{Configuration}. Apparently, the magnitude of the $\lambda(\bm{k})$ depends on the $\theta$. Therefore, the squeezing parameters depend on the direction of the wave-number vector of gravitons for the fixed magnetic field. 
For simplicity, we take $\theta =\pi/2$ in the following. But we consider the effects of the angle in Section~4.3.

In FIG.~\ref{graph1}, we plotted the squeezing parameter $\Lambda$ as a function of the scale factor $a(\eta)$. We see that the amplitude of $\Lambda$ goes to unity after the horizon exit and graviton and photon pair production become maximum during inflation. That is, the maximum entangled pairs of graviton and photon are produced. The FIG.~\ref{graph2} shows that graviton-photon pair production occurs but the production keeps decreasing after the horizon exit.

\noindent

\noindent

\section{Graviton state at present}

The standard lore about PGWs is 
that the quantum state of PGWs (gravitons) becomes squeezed during inflation due to the mechanism we discussed in the previous section.
Hence, it is believed that finding the squeezed gravitons turns out to prove inflation. 
In this section, we see the quantum state of gravitons 
can be a mixed state in the presence of magnetic fields.
This result tells us that we have to reconsider how to estimate the non-classicality of the primordial gravitational waves.

\subsection{Schmidt decomposition}

In the previous section, we found that the squeezing of the graviton-photon pair is produced but eventually disappears during inflation. In this subsection, we reveal the entanglement between the graviton-photon pairs. We compute the entanglement entropy between  gravitons and photons by tracing over the photons using the method developed in~\cite{Maldacena:2012xp,Kanno:2014lma}.

The initial state expressed in Eq.~(\ref{BD0}) and it is difficult to trace over the photon degree of freedom. Thus, we perform the following Bogoliubov transformation 
\begin{eqnarray}
\hat{C}_{y,{\bm k}}=\Phi\,\hat{a}_{y,{\bm k}}+\Psi\,\hat{a}_{y,-{\bm k}}^\dagger \ ,
\quad ~~
\hat{C}_{x,{\bm k}}=\Upsilon\,\hat{a}_{x,{\bm k}}+\Omega\,\hat{a}_{x,-{\bm k}}^\dagger \ ,
\label{transform}
\end{eqnarray}
where $|\Phi|^2-|\Psi|^2=1$, $|\Upsilon|^2-|\Omega|^2=1$ so that the state $|\overline{{\rm BD}}\rangle$ becomes in the Schmidt form
\begin{eqnarray}
|\overline{{\rm BD}}\rangle=\prod_{{\bm k}=-\infty}^\infty \exp\bigl[
\,\rho\,\hat{C}^\dagger_{y,{\bm k}}\,\hat{C}^\dagger_{x,-{\bm k}}
\bigr]
|0'\rangle_{y,{\bm k}}|0'\rangle_{x,-{\bm k}}\,.
\label{newBD}
\end{eqnarray}
Note that we consider different Bogoliubov coefficients between $(\Phi\,,\Psi)$ and $(\Upsilon\,,\Omega)$ because the $\Lambda$ and $\Xi$ in Eq.~(\ref{BD0}) are complex parameters. Here new vacuum states are defined by
\begin{eqnarray}
\hat{C}_{y,{\bm k}}\,|0'\rangle_{y,{\bm k}}=0
\ , \quad 
\hat{C}_{x,{\bm k}}\,|0'\rangle_{x,{\bm k}}=0\ .
\end{eqnarray}

Performing the new operators $\hat{C}_{y,{\bm k}}$ and $\hat{C}_{x,{\bm k}}$ on Eq.~(\ref{newBD}), we obtain the following relations,
\begin{eqnarray}
&&\hat{C}_{y,{\bm k}}\,
|\overline{{\rm BD}}\rangle
=\rho\,\hat{C}^\dagger_{x,-{\bm k}}\,
|\overline{{\rm BD}}\rangle,\\
&&\hat{C}_{x,{\bm k}}\,
|\overline{{\rm BD}}\rangle
=\rho\,\hat{C}^\dagger_{y,-{\bm k}}\,
|\overline{{\rm BD}}\rangle\,.
\end{eqnarray}
By using Eq.~(\ref{transform}), the above relations lead to the equations for the Bogoliubov coefficients as
\begin{eqnarray}
\begin{pmatrix}
\Lambda & 1& 0&-\rho\Xi\\
\Xi&0&-\rho&-\rho\Lambda \\
-\rho^*&-\rho^*\Lambda^*&\Xi^*&0\\
0&-\rho^*\Xi^*&\Lambda^*&1
\end{pmatrix} 
\begin{pmatrix}
\Phi\\
\Psi\\
\Upsilon^*\\
\Omega^*
\end{pmatrix} 
=0.
\label{newbogoliubov}
\end{eqnarray}
In order to find a nontrivial solution, the determinant of the above 4 by 4 matrix has to be zero. That is, $|\rho|^2$ satisfies
\begin{eqnarray}
|\rho|^2=Q
-\sqrt{Q^2-1}\,,
\end{eqnarray}
where we have defined
\begin{eqnarray}
Q =\frac{\left(|\Lambda|^2-1\right)^2+|\Xi|^4-2{\rm Re}(\Xi^2\Lambda^{*2})}{2|\Xi|^2}
\ .
\nonumber\\
\end{eqnarray}
\begin{figure}[H]
\centering
 \includegraphics[keepaspectratio,width=0.68\linewidth]{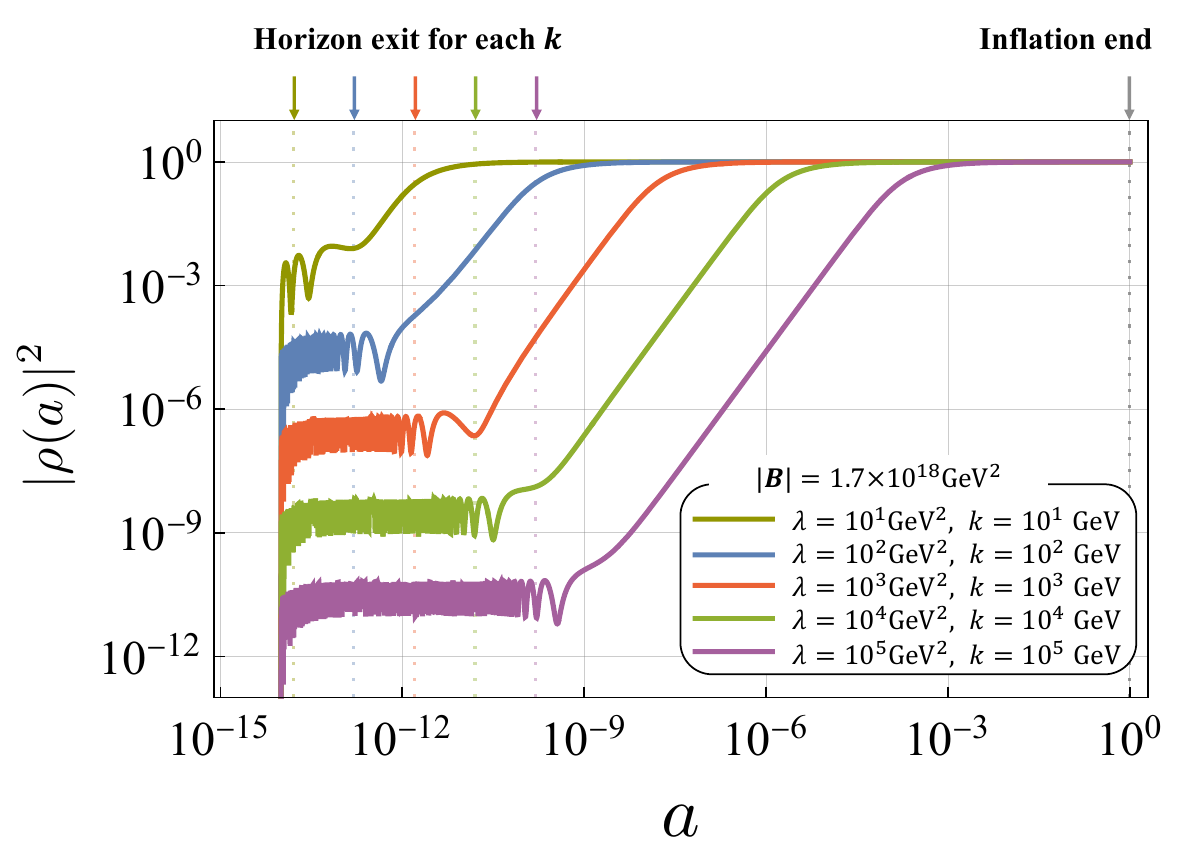}
 \renewcommand{\baselinestretch}{3}
 \caption{Plots of the parameter $|\rho (a)|^2$ as a function of $a(\eta)$.  Other parameters are set as $H=10^{14}\,{\rm GeV}, \eta_i=-1\,{\rm GeV}^{-1}, \eta_f=-10^{-14}\,{\rm GeV}^{-1},\lambda=10\,{\rm GeV}^2, a(\eta_i)= 10^{-14},a(\eta_f)=1$.}
 \label{graph3}
 \end{figure}
\noindent
In FIG.~\ref{graph3}, we plotted $|\rho|^2$ versus $a(\eta)$ for various values of $k$ under a fixed value of $|{\bm B}|$. Here, $\lambda$ is automatically determined once $k$ is fixed because of Eq.~(\ref{coupling2}) where we take $\theta=\pi/2$. We see that $|\rho|^2$ goes to unity irrespective of the value of $k$ after the horizon exit if the value of $|{\bm B}|$ is fixed.
Hence, the squeezing of graviton-photon pair in the basis $|0'\rangle_{y,{\bm k}}|0'\rangle_{x,-{\bm k}}$ turns out to be almost maximum, while $\Xi$ in the basis of $|0\rangle$ eventually vanishes as shown in FIG.~\ref{graph2}.

\subsection{Entanglement entropy}

Since gravitons and photons are coupled each other through $\lambda$ as in Eqs.~(\ref{eom:graviton}) and (\ref{eom:photon}), they are expected to get entangled eventually. In the previous subsection, we find the squeezing of graviton-photon pair becomes almost maximum in the basis of $|0'\rangle_{y,{\bm k}}|0'\rangle_{x,-{\bm k}}$ but eventually vanish in the basis of $|0\rangle$. In order to clarify whether they get entangled or not, we compute the entanglement entropy as a measure of entanglement. The entanglement entropy is basis independent. 

We define the density operator of the vacuum $|\overline{{\rm BD}}\rangle$ in Eq.~(\ref{newBD}) by 
\begin{eqnarray}
\sigma &=& |\overline{{\rm BD}}\rangle \langle \overline{{\rm BD}}|\nonumber\\
&=&\left(1-|\rho|^2\right)\prod_{{\bm k},-{\bm k}}\sum_{n',m'=0}^\infty\rho^{n'}\,\rho^{*m'}\,|n'\rangle_{y,{\bm k}}\,|n'\rangle_{x,{-\bm k}}\,\,
{}_{y,{\bm k}}\,\langle m'|\,{}_{x,-{\bm k}}\langle m'|
\ .
\end{eqnarray}
The reduced density operator
for the gravitons is obtained by tracing over the degree of freedom of photons such as
\begin{eqnarray}
\sigma_y &=& {\rm Tr}_x\,|\overline{{\rm BD}}\rangle \langle \overline{{\rm BD}}|=\sum_i{}_{x,{\bm k'}}\langle\,i\,|\overline{{\rm BD}}\rangle \langle \overline{{\rm BD}}|\,i\,\rangle_{x,{\bm k'}}\nonumber\\
&=&\left(1-|\rho|^2 \right)\sum_{n'=0}^\infty |\rho|^{2n}\, |n'\,\rangle_{y,{\bm k}}\,\,{}_{y,{\bm k}}\langle n'\,|
\label{graviton_state}\ .
\end{eqnarray}
The entanglement entropy between the graviton and photon can be characterized by
\begin{eqnarray}
S&=&-{\rm Tr}_y\,\sigma_y \log \sigma_y 
=-\sum_{n'=0}^\infty\left(1-|\rho|^2\right)|\rho|^{2n'}
\Bigl(\log\left(1-|\rho|^2\right)+n'\log|\rho|^2\Bigr)
\nonumber\\
&=&-\log \left(1-|\rho|^2\right)
  - \frac{|\rho|^2}{1-|\rho|^2}
  \log |\rho|^2 \ .
\end{eqnarray}

\begin{figure}[H]
\centering
 \includegraphics[keepaspectratio,width=0.68\linewidth]{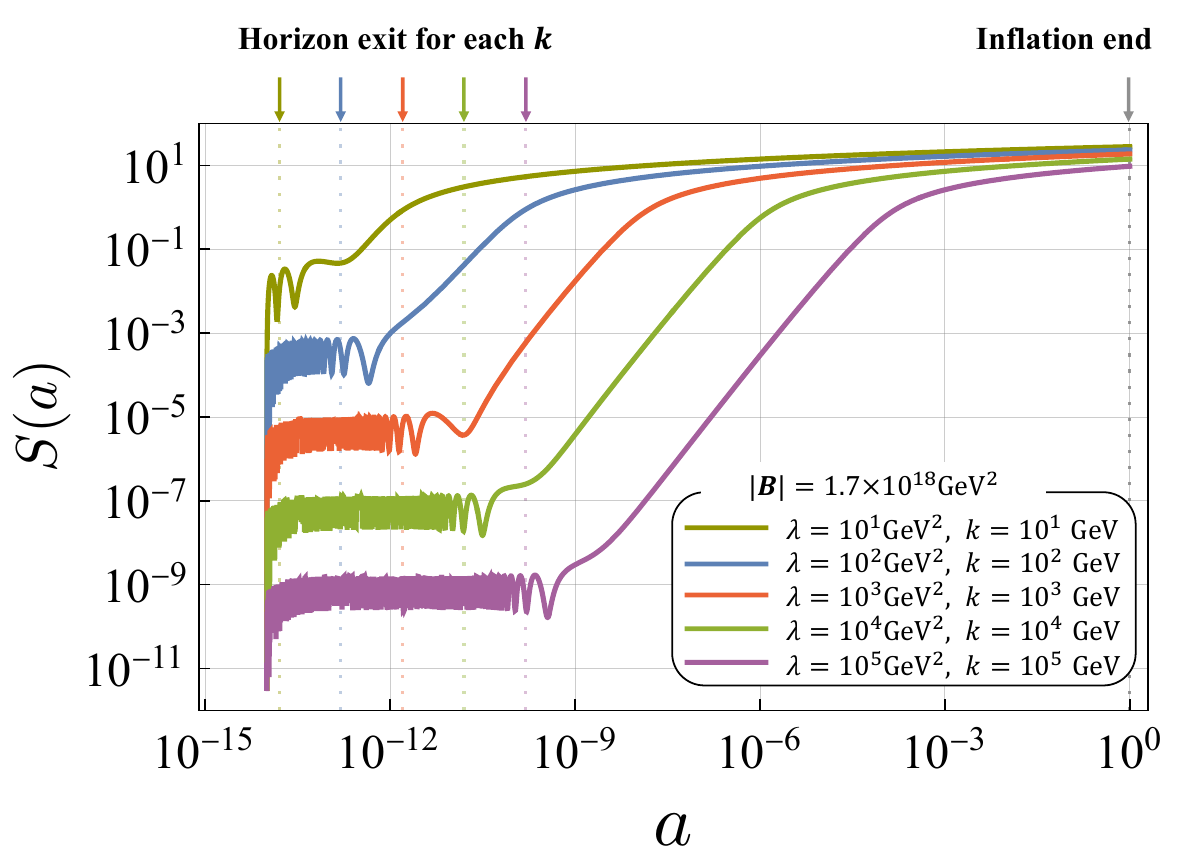}
 \renewcommand{\baselinestretch}{3}
 \caption{Entanglement entropy between graviton and photon as a function of $a(\eta)$. Other parameters are set as $H=10^{14}{\rm GeV}$, $\eta_i=-1{\rm GeV}^{-1}$, $\eta_f=-10^{-14}{\rm GeV}^{-1}$,  $a(\eta_i)= 10^{-14}$, and $a(\eta_f)=1$.}
 \label{graph4}
 \end{figure}
\noindent
In FIG.\,\ref{graph4}, we plotted the entanglement entropy for various values of $k$ under a fixed value of $|{\bm B}|$, which clearly shows that the graviton and photon are highly entangled during inflation. As well as the result of FIG.~\ref{graph3}, the asymptotic value of $S(a)$ becomes the same irrespective of the value of $k$.
\begin{figure}[H]
\centering\includegraphics[keepaspectratio,width=0.68\linewidth]{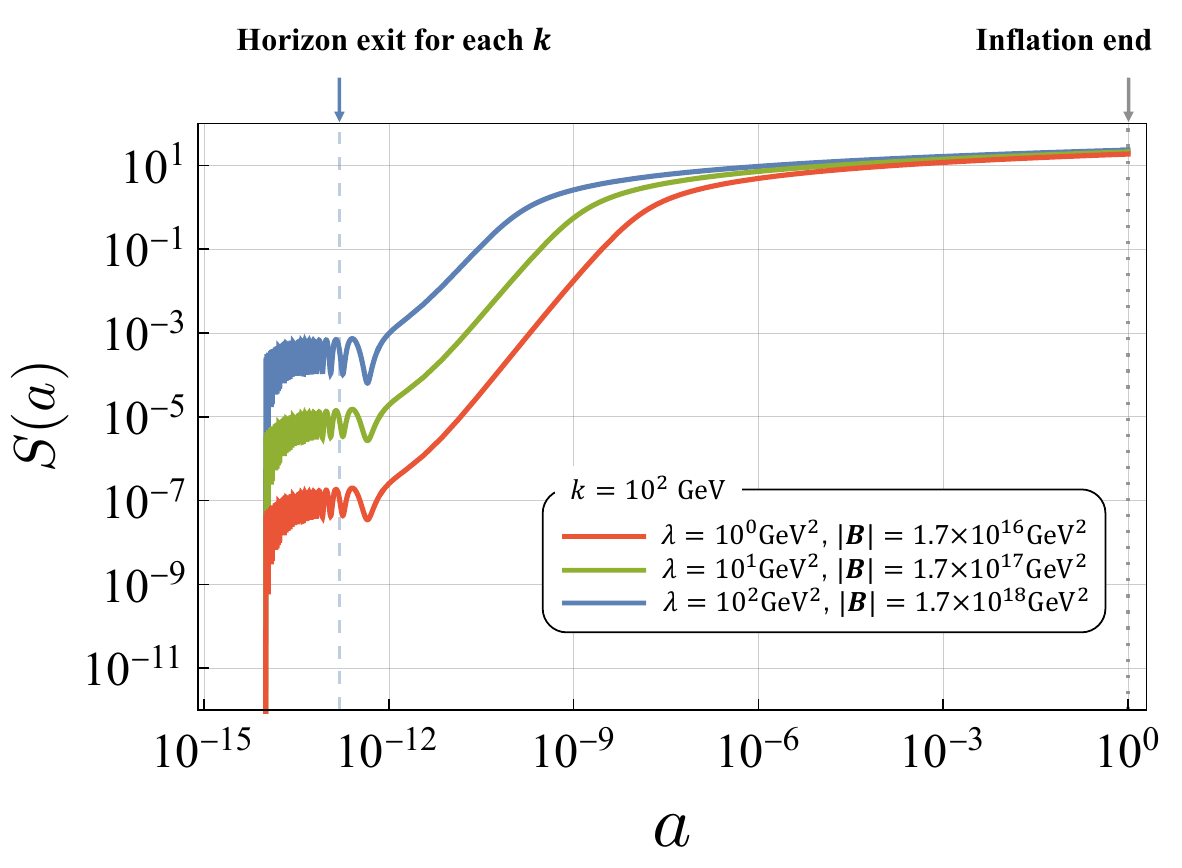}
 \renewcommand{\baselinestretch}{3}
 \caption{The entanglement entropy of graviton and photon field induced by the coupling of background magnetic field with a different magnetic field. Each line with a different color represents a different magnetic field. Other parameters are set as $H=10^{14}\,{\rm GeV}, \eta_i=-1\,{\rm GeV}^{-1},$ and $\eta_f=-10^{-14}\,{\rm GeV}^{-1}$, $a(\eta_i)= 10^{-14}$, and $a(\eta_f)=1$.}
 \label{graph5}
 \end{figure}
\noindent
In FIG.\,\ref{graph5}, the entanglement entropy for various values of $\lambda$ under a fixed value of $k$ is plotted. In this case, the different $\lambda$ corresponds to different $|{\bm B}|$ because of Eq.~(\ref{coupling2}) where $\theta=\pi/2$.

\section{Conclusion}
We studied primordial gravitational waves (PGWs) in the presence of magnetic fields that survive during inflation. 
In contrast to conventional inflation, where only PGWs are highly squeezed, we considered a system that electromagnetic fields are highly squeezed as well. 
We showed that graviton to photon conversion and vice versa never end as long as inflation lasts, and then gravitons and photons get highly entangled. 
We derived a reduced density matrix of the gravitons and calculated their entanglement entropy by using the reduced density matrix. 
We revealed that quantum states of the primordial gravitons observed today are not squeezed (pure) states but mixed states.

Our findings have important implications for the quantum state of primordial gravitons. 
So far, states of primordial gravitons are regarded as squeezed pure states. However, if magnetic fields had coupled with gravitons during inflation, the primordial gravitons observed today would be mixed states. 
Then the estimation of observables has to be changed.
Our results also open up the possibility of probing primordial magnetic fields through the observations of the non-classicality of primordial gravitational waves.
It is a challenge to witness such entanglement. However, it would be worth pursuing endeavors to find evidence of the quantum nature of primordial GWs.

As we mentioned below Eq.~(\ref{f}), we considered the coupling parameter $c=-1/2$ instead of the scale-invariant
one $c=-1$. 
In the case of $c=-1/2$, the physical magnetic fields decay as $fB\propto 1/a$, which is slower than the normal
scaling $\propto 1/a^2$. On the other hand, in the case of $c=-1$, the physical magnetic fields $fB$ does not decay during inflation. 
Hence, we would be able to expect more drastic effects on the quantum state of gravitons. 
We leave the analysis of this case for future work.

\section*{Acknowledgments}
S.\ K. was supported by the Japan Society for the Promotion of Science (JSPS) KAKENHI Grant Number JP22K03621.
J.\ S. was in part supported by JSPS KAKENHI Grant Numbers JP17H02894, JP17K18778, JP20H01902, JP22H01220.
K.\ U. was supported by the JSPS KAKENHI Grant Number 20J22946.

\printbibliography
\end{document}